\documentclass[twocolumn,aps,showpacs,prbprintnumbers,fleqn,superscriptaddress]{revtex4}
\usepackage{epsfig}
\usepackage{graphicx}
\usepackage{amsfonts}
\usepackage{subfigure}
\usepackage[dvips]{color}

\newcommand{\ct}{\cite}
\newcommand{\bi}{\bibitem}

\newcommand{\si}{\sigma}

\newcommand{\ket}{\rangle}
\newcommand{\non}{\nonumber}
\newcommand{\dg}{\dagger}
\newcommand{\be}{\begin{equation}}
\newcommand{\ee}{\end{equation}}
\newcommand{\ba}{\begin{eqnarray}}
\newcommand{\ea}{\end{eqnarray}}

\begin{document}

\title{Landau-Zener problem with waiting at the minimum gap and related
quench dynamics of a many-body system}
\author{Uma Divakaran}
\author{Amit Dutta}
\affiliation{Department of Physics, Indian Institute of Technology, Kanpur
208 016, India}
\author{Diptiman Sen}
\affiliation{Center for High Energy Physics, Indian Institute of Science,
Bangalore 560 012, India}

\date{\today}

\begin{abstract}
We discuss a technique for solving the Landau-Zener (LZ) problem of finding 
the probability of excitation in a two-level system. The idea of time 
reversal for the Schr\"odinger equation is employed to obtain 
the state reached at the final time and hence the excitation probability. 
Using this method, which can reproduce the well-known expression for the 
LZ transition probability, we solve a variant of the LZ 
problem which involves waiting at the minimum gap for a time $t_w$; we find 
an exact expression for the excitation probability as a function of $t_w$.
We provide numerical results to support our analytical expressions. We then 
discuss the problem of waiting at the quantum critical point of a many-body 
system and calculate the residual energy generated by the time-dependent 
Hamiltonian. Finally we discuss possible experimental realizations of this
work.
\end{abstract}

\maketitle

\section{Introduction}

Introduced in the 1930s independently by Landau and Zener \ct{landau,majorana},
the Landau-Zener (LZ) transition formula provides an exact expression for the 
excitation probability in the final state when two levels approach each other 
due to a linear variation of the diagonal terms of a two-level Hamiltonian 
with an avoided level crossing at the point of minimum gap. Even after 
seventy years, the LZ formula is being applied extensively in 
problems over a wide range of modern physics, including neutrino oscillations,
atomic physics, quantum optics, and mesoscopic systems \ct{lz_applications}. 
Recently, it has found several applications in quantum computations 
\ct{johansson08} and adiabatic quantum dynamics of many-body systems 
\ct{santoro02}. In particular, the general Kibble-Zurek (KZ) scaling 
\ct{kibble76,zurek96,polkovnikov05,dziarmaga09} of the residual energy or the 
defect density produced in the final state of a many-body Hamiltonian 
following a slow passage across a quantum critical point (QCP) \ct{sachdev99} 
has been established for a number of exactly solvable low-dimensional 
non-random spin 
models using the LZ formula \ct{dziarmaga05,levitov06,sengupta08,sen08}. 
All these systems factorize into decoupled $2 \times 2$ matrices in momentum 
space, and the dynamics gets mapped to a set of decoupled LZ problems.
We note that there are several other studies of quenching dynamics of 
different critical systems \ct{cazalilla06, rossini09,polkovnikov09}. 
Quenching through a multicritical point \ct{divakaran091}, 
repeated passage through a QCP \ct{mukherjee08} and periodic variation
of a parameter \ct{mukherjee09}, quenching dynamics of a disordered spin 
chain \ct{dziarmaga06}, and the dynamics of an open system coupled to a heat
bath \ct{patane08} have also been studied.

In this paper, we address the following question: how does the transition 
probability of a LZ problem get altered if we introduce a waiting time 
$t_w$ at the minimum gap? Here, starting from the ground state at $t = -
\infty$, the system is brought to the point of minimum gap ($2\Delta$) at time
$t=0$ by linearly varying the diagonal terms of the Hamiltonian at a rate 
$1/(2\tau)$. At the minimum gap, the system is allowed to evolve without any 
external driving for a time $t_w$, after which the linear variation is again 
resumed up to $t = \infty$. We study how the probability of finding the system 
in the excited state at $t = \infty$, given exactly by $\exp(-2 \pi 
\Delta^2 \tau)$ for the conventional LZ problem \ct{landau},
gets modified due to the additional time scale $t_w$. We then
extend our results for the Landau-Zener problem with waiting
to a many-body system whose quench dynamics through a QCP can be viewed as 
a set of decoupled LZ problems; the system is brought from its 
initial ground state at $t = -\infty$ to the QCP at $t=0$, where the 
Hamiltonian is not changed for a time $t_w$, before it is again varied in 
time up to $t = \infty$. As the relaxation time of the system diverges
at the critical point, the response of the system to any perturbation
becomes infinitely slow; hence the system is no longer able to follow the 
ground state and therefore excitations are produced. As mentioned 
already, there have been several studies in this field recently,
but the effect of waiting for a time $t_w$ at the critical point of a
many-body system has not been studied. In a many-body system, it is more 
meaningful to look at quantities like the residual energy $e_r$ defined as 
the difference in energy between the actual state reached and the true 
ground state at the final time, or the density of defects (wrongly oriented 
spins) $n$ which is obtained by integrating the probability of excitations 
over all modes.

We will study the effect of waiting on the residual energy $e_r$ in 
the final state of the one-dimensional Kitaev model following a quench through 
the QCP along with waiting. We study the possible correction to 
the KZ power law scaling $n\sim 1/\tau^{d \nu/(\nu z+1)}$, where $d$ is 
the spatial dimension and $\nu$ and $z$ are the correlation 
length and dynamical exponents associated with the QCP \ct{sachdev99} 
across which the system is swept. To the best of our knowledge, these 
questions have not been addressed before from a 
theoretical point of view, although experimental results with waiting
for single molecular magnets ${\rm Mn_{12}Ac}$ are available \ct{hill06}. 
The possibility of experimental realizations of quenching dynamics with 
waiting in optical lattices serves as another motivation. It is interesting 
to note that a similar concept is used in Ramsey spectroscopy where a molecule
is subjected to an oscillating perturbation for a time $T_1$ which induces 
transitions between two specific levels of the molecule. The perturbation is 
then switched off for a time $T_2$ after which it is again switched on for a 
time $T_1$ and the probability of excitations is found \ct{ramsey}. We 
note that LZ sweeps have been used to generate coherent superpositions in 
quantum optical experiments \ct{yatsenko02}.

Our results can be summarized as follows. The effect of a waiting time $t_w$ 
can be understood by visualizing the dynamics in two parts: from $t = -\infty$
to $0$ and from $t=0$ to $\infty$. If the solution of the first part is known,
the second part can be solved by applying the idea of time 
reversal on the first part; in this way, given the amplitudes of the two 
basis states at $t=0$, one can find the values of the same at $t = \infty$. 
This reproduces the exact result for the conventional LZ problem. We then 
apply the method to a problem with waiting at the minimum gap to obtain an 
exact expression for the excitation probability in the final state; this gives
simple forms for both the diabatic ($\Delta^2 \tau \to 0)$ and the 
adiabatic ($\Delta^2 \tau \to \infty$) limit. The probability of excitations
exhibits a sinusoidal behavior in both cases but with different pre-factors 
and phases. The method is then used for a many-body system, namely, the 
one-dimensional Kitaev model \ct{kitaev06} to obtain the residual energy. We 
quench this system through a QCP by linearly varying the anisotropy in the 
interaction, $dJ_- /dt = 1/\tau$ \ct{sengupta08,divakaran09}, with a waiting 
time $t_w$ at the QCP. We show that for $t_w/\sqrt{\pi \tau} \ll 1$, the 
residual energy shows an exponential decay with $t_w$ given by $e_r \sim 
(a/{\sqrt \tau}) [1 + b \exp(-ct_w/{\sqrt \tau})]$. We find that the 
parameters $a$, $b$ and $c$ obtained by an approximate analytical 
calculation give a good fit with the results obtained by numerically 
solving the Schr\"odinger equation.
 
The outline of this paper is as follows. We describe our method of solving 
the Landau-Zener method in Sec. \ref{sec_lz} and obtain an expression 
for the excitation probability in the presence of waiting in Sec. 
\ref{sec_lzw}. In Sec. \ref{sec_kitaev}, we study the waiting problem for 
a many-body system taking the example of the one-dimensional Kitaev model.

\section{Landau-Zener problem}

\subsection{Landau-Zener revisited}\label{sec_lz}

To illustrate our method, let us revisit the conventional two-level LZ 
problem. Although the model has been exactly solved and the evolution matrix 
is known exactly for all times \ct{vitanov96,kayanuma}, and it has also been 
studied within a rotating wave approximation \ct{wubs05}, we describe a method
below which can be easily generalized to study LZ dynamics with waiting.

The state $\psi (t) = C_1(t)|1\ket + C_2(t)|2\ket$, 
where $|1\ket$ and $|2\ket$ are the basis states (the initial and final 
ground states, respectively), evolves according to the Schr\"odinger equation
\ba i \frac{d}{dt} \psi(t) ~=~ \left[ \begin{array}{cc} 
t/(2\tau) & \Delta \\
\Delta & -t/(2\tau) \end{array} \right] \psi(t) ~=~ H \psi(t),
\label{gen_mat} \ea
where $\Delta$ is chosen to be real without any loss of generality. With the 
initial condition $|C_1(-\infty)|^2=1$, the wave function at $t=0$ is given by
$\psi(t=0) = \alpha |1\ket + \beta |2\ket$, where
\ba \alpha &=& C_1 (0) ~=~ e^{-\frac{\pi}{4}\Delta^2 \tau}~
e^{i\frac{3\pi}{4}}~ \sqrt{\pi} ~\frac{2^{-iy}}{\Gamma(1/2+iy)}, \non \\
\beta &=& C_2 (0) ~=~ \Delta\sqrt{\tau} ~e^{-\frac{\pi}{4}\Delta^2\tau}
\sqrt{\frac{\pi}{2}}~ \frac{2^{-iy}}{\Gamma(1+iy)}, \ea
with $y=\Delta^2 \tau/2$ \ct{vitanov96,damski06}. Henceforth, 
$(\cdot , \cdot)^T$ will denote the transpose of a given row vector. Since 
$\psi(-\infty) = (1,0)^T$ evolves to $\psi(0) = (\alpha, \beta)^T$, 
orthogonality implies that $\psi(-\infty) = (0,1)^T$ must evolve to 
$\psi(0) = (\beta^*, -\alpha^*)^T$, up to a phase. Using properties of the 
Gamma functions \ct{abram}, one can show that $|\alpha|^2 + |\beta|^2 = 1$ 
as desired, and $|\alpha|^2 - |\beta|^2 ~=~ e^{-\pi\Delta^2\tau}$.

Let us now ask: what wave functions at $t=0$ will evolve to $(1,0)^T$ and 
$(0,1)^T$ at $t = \infty$? To answer this question, let us multiply the
Hamiltonian in Eq. (\ref{gen_mat}) by $\sigma^z$ on both sides, which gives
\ba -i \frac{d}{dt} \si^z \psi (t) ~=~ \left[ \begin{array}{cc} 
- t/(2\tau) & \Delta \\
\Delta & t/(2\tau) \end{array} \right] \si^z \psi (t). \label{time_mat} \ea
Clearly, by substituting $t'=-t$ and $\psi' (t')= \si^z \psi (t)$
in Eq. (\ref{time_mat}), we recover Eq. (\ref{gen_mat}). 
Thus, the dynamics occurring in Eq. (\ref{gen_mat}) from
$t= 0$ to $- \infty$ is the same as in Eq. (\ref{time_mat}) with $t'$ 
going from $0$ to $\infty$ and $\psi$ replaced by $\si^z \psi$. 
Since $\psi(-\infty) = (1,0)^T$ evolves to $\psi(0) = (\alpha, \beta)^T$, 
the above argument shows that $\psi(0)= \si^z (\alpha, \beta)^T = (\alpha, 
-\beta)^T$ evolves to $\psi(\infty) = (1,0)^T$ through the Hamiltonian given 
in Eq. (\ref{gen_mat}). By similar arguments, or by orthogonality, we see that 
$\psi(0)= \si^z (\beta^*, -\alpha^*)^T = (\beta^*, \alpha^*)^T$ evolves to
$\psi(\infty) = (0,1)^T$, again up to a phase. We can now find the probability
of ending in the excited state $|1\ket$ at $t=\infty$ as follows. We can write
$\psi(0) = (\alpha, \beta)^T$ as
\ba \psi(0) &=& \left[ \left( \begin{array}{c} 
\alpha \\ 
-\beta \end{array} \right) (\alpha^*, -\beta^*) ~+~ \left( \begin{array}{c} 
\beta^* \\ 
\alpha^* \end{array} \right) (\beta, \alpha) \right] \left( \begin{array}{c}
\alpha \\ 
\beta \end{array} \right ) \non \\ 
&=& (|\alpha|^2 - |\beta|^2) ~\left( \begin{array}{c} 
\alpha \\ 
-\beta \end{array} \right) + 2 \alpha \beta \left( \begin{array}{c} 
\beta^* \\ 
\alpha^* \end{array} \right) \label{identity3} \ea
where, in the first line, we introduced an identity operator using 
the orthonormal basis $(\alpha,-\beta)^T$ and $(\beta^*,\alpha^*)^T$. 
Since we now know the evolution of $(\alpha,-\beta)^T$ and 
$(\beta^*,\alpha^*)^T$ from $t=0$ to $\infty$, we see that
\ba \psi(\infty)= (|\alpha|^2 - |\beta|^2) ~\left( \begin{array}{c} 
1 \\ 
0 \end{array} \right) ~+~ 2 \alpha \beta~ \left( \begin{array}{c} 
0 \\ 
1 \end{array} \right), \ea
where there may be an unimportant phase difference between the two terms.
The probability of excitations at the final time is the probability
to be in state $|1\ket$ and is thus given by
$p = \left( |\alpha|^2 - |\beta|^2 \right)^2 = e^{-2\pi\Delta^2\tau}$
which is the exact expression for the LZ transition probability \ct{landau}.

\subsection{Landau-Zener with waiting}\label{sec_lzw}

We now apply our method to the LZ problem with waiting. 
Within the time interval $[0,t_w]$, the eigenvectors of $H$ 
are given by $(1,\pm 1)^T /\sqrt{2}$ with eigenvalues $\pm \Delta$.
So the wave function changes from $\psi (0) = (\alpha,\beta)^T$ to
\ba \psi(t_w) = \frac{\alpha+\beta}{2} ~e^{-i\Delta t_w} \left( 
\begin{array}{c} 1 \\ 
1 \end{array} \right ) + \frac{\alpha-\beta}{2} ~e^{i\Delta t_w} \left( 
\begin{array}{c} 1 \\ 
-1 \end{array} \right ) \non \ea
Inserting the identity operator used in Eq. (\ref{identity3}), we get
\ba \psi(t_w) = \left[ \left( \begin{array}{c} \alpha \\ 
-\beta \end{array} \right) (\alpha^*, -\beta^*) + \left( \begin{array}{c} 
\beta^* \\ 
\alpha^* \end{array} \right ) (\beta, \alpha) \right] \psi(t_w). \non \ea
The state at $t = \infty$ can again be obtained by using the information
about the evolution of $(\alpha,-\beta)^T$ and $(\beta^*,\alpha^*)^T$:
\ba \psi(\infty)= (\alpha^*, -\beta^*)\psi(t_w) \left( \begin{array}{c} 1 \\ 
0 \end{array} \right) ~+~ (\beta, \alpha)\psi(t_w) \left( \begin{array}{c} 0 \\
1 \end{array} \right). \non \ea
Hence the excitation probability is given by
\ba p_{t_w} &=& |(\alpha^*, -\beta^*)\psi(t_w)|^2 \non \\
&=& \left[ (|\alpha|^2-|\beta^2|)\cos(\Delta t_w) -i (\alpha^*\beta-\alpha
\beta^*)\sin(\Delta t_w) \right]^2 . \non \\ \label{analytical} \ea
This expression simplifies in the limits 
$\Delta^2 \tau \to 0$ and $\infty$. For $\Delta^2 \tau = 0$, we have $p_{t_w}
=\cos^2(\Delta t_w)$; this is expected as the system does not get any 
time to evolve and therefore remains in the state $|1\ket$ up to $t=0$, then 
oscillates between the states $|1\ket$ and $|2\ket$ from $t=0$ to $t_w$, and
then remains in the superposed state reached at $t_w$ for all $t>t_w$. If 
$\Delta^2 \tau \ll 1$, one gets $i(\alpha^*\beta - \beta^* \alpha) ~\simeq~ 
\Delta \sqrt{\pi \tau}$ which leads to an approximate expression
\be p_{t_w} ~\simeq~ e^{-\pi \Delta^2 \tau} ~\cos^2 [\Delta (t_w + 
\sqrt{\pi \tau})]. \label{approx1} \ee
Comparison with the case $\Delta^2 \tau = 0$ shows a decrease in amplitude 
from 1 to $e^{- \pi \Delta^2 \tau}$ and a phase shift of $\sqrt{\pi \tau}$. In 
the adiabatic limit $\Delta^2 \tau \to \infty$, $|\alpha|^2-|\beta|^2=0$, and 
one can use the asymptotic expansions of the Gamma function \ct{abram}
to obtain the excitation probability 
\ba p_{t_w} ~\simeq~ \frac{1}{16\Delta^4 \tau^2} ~\sin^2(\Delta t_w). \ea
In the next section, we consider waiting in a many-body problem, namely, the
one-dimensional Kitaev model.

\section{Waiting in Kitaev model}\label{sec_kitaev}

We now use the above results to study the quenching dynamics with waiting at 
the QCP of the one-dimensional Kitaev model given by the Hamiltonian
\ct{kitaev06,divakaran09}
\ba H ~=~ \sum_j ~(J_1 \si^x_{2j} \si^x_{2j+1} ~+~J_2\si^y_{2j-1} \si^y_{2j}),
\label{kitaev1} \ea
where $j$ refers to the site index and $\si^{x,y}$ are the Pauli matrices. 
The model can be exactly solved in terms of Jordan-Wigner fermions \ct{lieb61}
defined as
\ba a_j ~=~ (\prod_{i=-\infty}^{2j-1} \si_i^z) ~\si_{2j}^y,~~~~
b_j ~=~ (\prod_{i=-\infty}^{2j} \si_i^z) ~\si_{2j+1}^x. \ea
Going to momentum space with $\psi_k \equiv (a_k,b_k)^T$, where $a_k$ and 
$b_k$ are the Fourier transform of $a_j$ and $b_j$, and performing 
an appropriate unitary transformation, the Hamiltonian
decouples into a $2 \times 2$ form given by \ct{divakaran09}
\ba H &=& \sum_{k=0}^{\pi/2} ~\psi_k^\dg ~H_k ~\psi_k, \non \\
{\rm where}~~~ H_k &=& 2 \left[ \begin{array}{cc} J_- \sin(k) & J_+\cos(k) \\
J_+ \cos(k) & -J_- \sin(k) \end{array} \right], \label{kitaev2} \ea
$J_\pm = J_1 \pm J_2$ and $k$ ranges from $0$ to $\pi/2$.
The vanishing of the gap ($=4 \sqrt{J_-^2 \sin^2 k + J_+^2 \cos^2 k}$) for 
the mode $k=\pi/2$ at $J_- =0$ signals a quantum phase transition of 
topological nature \ct{feng07}, with $\nu = z = 1$.

\begin{figure} \includegraphics[height=3.1in,angle=-90]{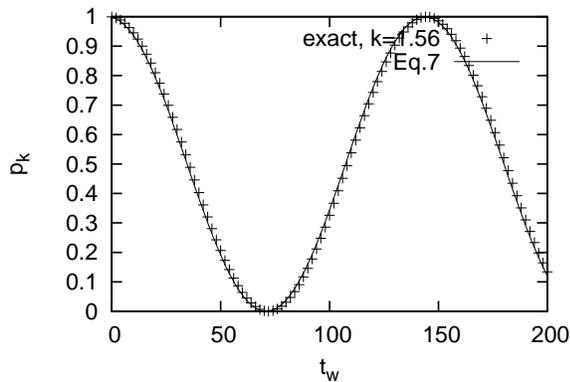}
\caption{Variation of probability of excitations $p_k$ with waiting time 
$t_w$, for $\tau=1$ and $k=1.56$. The dots correspond to the exact solution 
given in Eq. (\ref{analytical}), whereas the line corresponds to the 
approximate analytical expression given in Eq. (\ref{approx1}).}
\label{wait_pkvstw} \end{figure}

Setting $J_+ = 1$, we now apply the quenching scheme 
\ba J_- (t) &=& t/\tau ~~ {\rm for} ~~-\infty <t \le 0 \non \\
&=& 0~~ {\rm for} ~~0 \le t \le t_w \non \\
&=& (t - t_w)/\tau ~~ {\rm for} ~~t_w \le t < \infty, \label{quenching} \ea
which incorporates waiting for a time $t_w$ at the QCP. The Schr\"odinger 
equation for each mode is 
\ba i \frac{d}{dt} \psi_k(t) ~=~ 2 \left[ \begin{array}{cc} 
J_- (t) \sin k & \cos k \\
\cos k & - J_- (t) \sin k \end{array} \right] \psi_k(t),
\label{schrodinger} \ea
so that the excitation probability for each mode is given by the modified LZ
formula in Eq. (\ref{analytical}). In Fig. \ref{wait_pkvstw}, we compare the 
exact analytical expression given in Eq. (\ref{analytical}) with its 
corresponding approximate form for small $\Delta^2 \tau$ given in Eq. 
(\ref{approx1}); the picture in the figure is reminiscent of Ramsey fringes. 
Noting that $\Delta^2 \tau = \tau \cos^2 k /\sin k \simeq 0.00012$ for the 
mode shown in Fig. \ref{wait_pkvstw}, we can use the diabatic limit result in 
Eq. (\ref{approx1}), 
\be p_{k,t_w} \simeq e^{-\pi \tau \cos^2 k /\sin k} ~
\cos^2 [2 \cos k (t_w + \frac{1}{2} \sqrt{\frac{\pi \tau}{\sin k}})]. 
\label{approx2} \ee
This gives a good understanding of the peak heights and phase shift in 
Fig. \ref{wait_pkvstw}. 

\begin{figure} \includegraphics[height=3.1in,angle=-90]{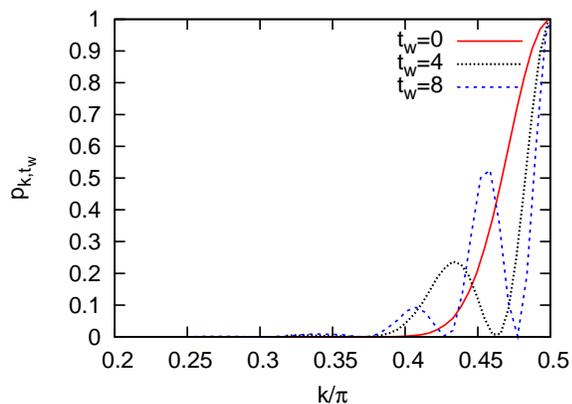}
\caption{(Color online) Variation of probability of excitations $p_{k,t_w}$ 
with $k$ for $\tau=10$ obtained by numerically solving the Schr\"odinger 
equation, for $t_w =0$ (solid red line), $4$ (dotted black line) and $8$ 
(dashed blue line).} \label{wait_pkvsk} \end{figure}

The variation of $p_{k,t_w}$ vs $k$ for different $t_w$ (shown in Fig.
\ref{wait_pkvsk} with $\tau=10$ ) shows secondary maxima, the peak heights of 
which increase as $t_w$ increases, in contrast to the conventional quenching 
case \ct{divakaran09}. The increase in the number of maxima with increasing 
$t_w$ can be explained by the expression in Eq. (\ref{approx2}) which has 
maxima at
\be \cos k ~=~ \frac{m \pi}{2(t_w + \sqrt{\pi \tau}/2)}, ~~~m=0,1,2,\cdots 
\label{cosk} \ee
in the limit of small $k$. With increasing $t_w$, Eq. (\ref{cosk}) is 
satisfied by more and more values of $k$ which still satisfy the condition
that $\cos k$ is small enough so that the peak height given by the exponential 
pre-factor in Eq. (\ref{approx2}) is not very small. Further, for a given 
value of $m$ in Eq. (\ref{cosk}), $\cos k$ decreases as $t_w$ increases which
justifies the increase in the peak height given by Eq. (\ref{approx2}).

\begin{figure} \includegraphics[height=3.1in,angle=-90]{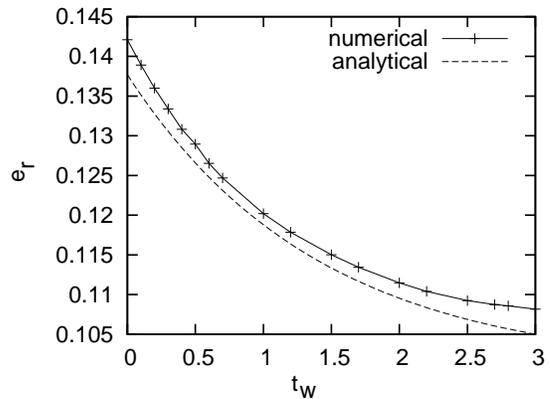}
\caption{Variation of residual energy $e_r$ with waiting time $t_w$, for 
$\tau = 10$. The line corresponds to the integral given in Eq. (\ref{resen}),
while the dotted line corresponds to Eq. (\ref{approx4}).} 
\label{wait_nvstw} \end{figure}

The residual energy per site, $e_r$, is given by the difference of the 
expectation value of the operator 
\be {\cal O} ~=~ \frac{1}{N} ~\sum_m ~(\si^x_{2m} \si^x_{2m+1} ~-~ 
\si^y_{2m-1} \si^y_{2m}) \ee
between the many-body state that is actually reached and the true ground state
of $H$ at $t = \infty$ (this is also the ground state of $\cal O$); $N$
denotes the number of sites. We find that $e_r$ is given by 
\be e_r ~=~ \int_{0}^{\pi/2} \frac{dk}{2\pi} ~8 \sin k ~p_{k,t_w}. 
\label{resen} \ee
Although this expression cannot be evaluated analytically in general, one can 
obtain an approximate expression when $\tau \gg 1$ and $t_w/\sqrt{\pi \tau} \ll
1$. Most of the contribution to the integral in Eq. (\ref{resen}) then comes 
from $k$ close to $\pi/2$ in Eq. (\ref{approx2}). Approximating $\cos k \simeq 
\pi/2 - k$ and $\sin k \simeq 1$, redefining $k = \pi /2 - k$ and finally 
extending the limits of integration from $[0,\pi/2]$ to $[0,\infty]$, we 
obtain
\ba e_r &=& \frac{2}{\pi}~ \int_0^\infty ~dk ~e^{-\pi \tau k^2} ~\left[ 1 ~+~
\cos [ 4 k (t_w + \frac{1}{2} \sqrt{\pi \tau})] \right] \non \\
&=& \frac{1}{\pi \sqrt{\tau}} ~[~ 1 ~+~ e^{-4 (t_w + \frac{1}{2} 
\sqrt{\pi \tau})^2/(\pi \tau)}~] \non \\
&\simeq& \frac{0.32}{\sqrt{\tau}} ~[~ 1 ~+~ 0.37 ~e^{-2.3 t_w/\sqrt{\tau}}~],
\label{approx4} \ea
where we have used the approximation $t_w/\sqrt{\pi \tau} \ll 1$ in the third 
line. In Fig. \ref{wait_nvstw}, the numerical results obtained 
by solving the Schr\"odinger equation are compared with the approximate 
analytical expression given in Eq. (\ref{approx4}). This comparison shows 
that the numerical and analytical expressions are in good agreement.

The decrease in the residual energy as a consequence of 
waiting can be understood as follows. For slow driving, only modes close to 
the critical modes contribute. For a mode with momentum $k$, the frequency 
of oscillations of $|C_1 (t)|^2$ between the two levels during waiting is 
proportional to $\Delta_k=2 \cos k$; this vanishes as $k$ approaches the
critical value $\pi/2$, leading to a diverging time period $T$. Further, 
the vanishing of the off-diagonal element in the Hamiltonian in 
Eq. (\ref{schrodinger}) implies that the modes close to $k=\pi/2$ will 
remain close to the excited state under time evolution, i.e., $|C_1|$ will 
remain close to 1. Now consider the variation of the diabatic excitation 
probability $|C_1|^2$ between $t=0$ and $t=t_w$. The waiting gives $|C_1|$ 
the time to oscillate to $|C_2|$. Since we are considering only values of 
$t_w$ much smaller than the time period $T$, we encounter only the decreasing 
part of the oscillating $|C_1|^2$ such that when the variation of the 
parameter is again started at $t_w$, $|C_1(t_w)|^2 < |C_1(t=0)|^2$. This 
reduces the probability of excitations at the final time.

The advantage of studying the waiting problem in the Kitaev model is that the 
minimum gap for all the modes occurs at the same time ($t=0$) which makes the
analytical calculations easier. In many other models, the minimum gaps for 
the different modes, given by the vanishing of the diagonal term of the $2 
\times 2$ Hamiltonians, do not occur at the same time. This makes it 
difficult to specify the precise time at which 
the waiting should be initiated so that our analytical results 
can be applied. But one can prove that in the limit of large $\tau$, the 
minima for all the modes approach the time at which the minima of the 
critical mode occurs. For large $\tau$, we therefore expect our analysis to 
go through with small corrections to the probability of excitations given 
in Eq. (\ref{approx2}). To understand why this is so, consider the quenching 
of the transverse magnetic field 
$h$ in the transverse field $XY$ model \ct{levitov06}, where the 
diagonal term of the equivalent $2 \times 2$ matrix is $h+J\cos k$ and the
off-diagonal term is $\gamma \sin k$, with $J=J_x+J_y$ and $\gamma=J_x-J_y$. 
By expanding the diagonal term about the critical mode
$k=0$, multiplying the corresponding Schr\"odinger equation with 
$\sqrt{\tau}$ and redefining $t'=t/\sqrt{\tau}$, the diagonal term can be 
rewritten as $t'+1/(2\gamma\sqrt{\tau})$, where the characteristic momentum 
scale is given by $1/(\gamma\sqrt{\tau})$ by the LZ tunneling formula.
Clearly, in the limit $\tau \to \infty$, the minima for the mode occurs
approximately at $t'=0$. These arguments are applicable to the modes close 
to the critical mode; the modes far away from the critical mode anyway do 
not contribute to the residual energy or defect density as the excitation 
gap is very large.

The waiting problem is also interesting from an experimental point of view. 
In Ref. \onlinecite{hill06}, the relaxation dynamics with waiting at 
a resonance was studied for single molecular magnets (SMM) called 
${\rm Mn_{12}Ac}$, where each molecule has total spin $S=10$. As the magnetic 
field in the $\hat z$-direction is varied from a large negative to a large
positive value, $S^z$ changes from $10$ to $-10$. In Ref. \ct{hill06}, the 
magnetic field is swept to a resonance value, starting from a field of 
$-6 ~T$, where it is held for different waiting times causing tunneling of 
the spins; eventually the field is brought back to its initial value. The 
number of molecules which have tunneled through the barrier, measured using 
electron paramagnetic resonance techniques, shows nearly a stretched 
exponential decay with $t_w$, where the decay constant is governed by 
the relaxation time of the system. Though there are limitations
in mapping SMM to a LZ problem for all quenching rates \ct{wernsdorfer05}, 
it is worth noting that there have been experimental studies on the effect 
of waiting at the resonance, and important quantities 
like the relaxation time of the system can be obtained from the
waiting problem. Although, in our case, $J_-$ is driven to a final 
value of $+\infty$ at $t =+ \infty$, we do observe a qualitatively 
similar behavior, i.e., an increase in the tunneling probability to 
the second state with increasing waiting time. We believe
that similar experiments with waiting and forward driving of the magnetic
field can be performed and it would be interesting to compare the results
with our predictions. However, if the non-linear term of a SMM Hamiltonian
\ct{wernsdorfer05} dominates, one may need to look at non-linear LZ problems 
\ct{nonlinearLZ} with waiting.

It may be mentioned here that quenching with waiting can be studied if the 
Kitaev model can be experimentally realized using cold atoms and molecules 
trapped in an optical lattice as proposed in Ref. \onlinecite{duan03}. In this
proposal, each of the couplings can be independently tuned using different 
microwave radiations. It is possible to investigate the evolution of the 
spatial correlation function of the operator $ib_n a_{n+r}$, defined in Ref. 
\onlinecite{sengupta08} as a function of various parameters, where $a_n$ and 
$b_n$ denote some Majorana fermions operators. This spatial correlation 
function depends on $p_k$ which we have already obtained for the waiting case.
Then the evolution of defect correlations can be detected by spatial noise 
correlation measurements as discussed in Ref. \onlinecite{altman04}.

\section{Conclusion}

To summarize, the technique proposed here not only provides an exact 
result for the standard LZ problem but also enables us to estimate the 
excitation probability for the dynamics with waiting at the minimum gap 
and the residual energy for the related quenching dynamics of 
some many-body systems like the one-dimensional Kitaev model. We can derive 
simple expressions in some limiting situations and the approximate analytical 
results are in excellent agreement with numerical results when $t_w/
\sqrt{\pi \tau} <<1$. The arguments leading up to Eq. (\ref{approx4}) indicate
that the KZ scaling law $1/\tau^{d\nu/(\nu z +1)}$ will generally remain 
valid in the presence of waiting, except that the function multiplying the 
scaling term has a piece which decays with increasing $t_w$. Finally, 
we have discussed some possibilities for experimentally testing our results.

\section{Acknowledgments}

U.D. and A.D. acknowledge T. Caneva, A. Polkovnikov, D. Rossini and G. E. 
Santoro for interesting discussions. U.D. also thanks S. Hill for a private 
communication. U.D. acknowledges CHEP in the Indian Institute of Science, 
Bangalore for its hospitality where part of this work was done.

\end{document}